\documentclass{emulateapj}
\usepackage{graphicx}
\usepackage{apjfonts}

\newcommand{\lsun}{\mbox{L$_{\sun}$ }}
\newcommand{\lsunend}{\mbox{L$_{\sun}$}}
\newcommand{\lir}{\mbox{L$_{\rm IR}$}}

\newcommand{\lmol}{\mbox{L$_{\rm mol}$}}

\newcommand{\lcojone}{\mbox{CO (J=1-0)}}
\newcommand{\lcojthree}{\mbox{CO (J=3-2)}}
\newcommand{\ncrit}{\mbox{$n_{\rm crit}$}}

\newcommand{\cmthree}{\mbox{cm$^{-3}$}}

\newcommand{\htwo}{\mbox{H$_2$}}

\shorttitle{Dense Molecular Gas in Galaxy Nuclei}

\shortauthors{Narayanan et al.}

\slugcomment{ApJL - accepted}

\begin{document}
\title{The Star Formation Rate - Dense Gas Relation in the Nuclei of
  Nearby Galaxies} \author{Desika Narayanan\altaffilmark{1,2}, Thomas
  J. Cox\altaffilmark{1,3}, and Lars Hernquist\altaffilmark{1}}
\altaffiltext{1}{Harvard-Smithsonian Center for Astrophysics, 60 Garden Street,
  Cambridge, MA 02138, USA}
\altaffiltext{2}{dnarayanan@cfa.harvard.edu}
\altaffiltext{3}{W.M. Keck Postdoctoral Fellow}

\begin{abstract}

We investigate the relationship between the star formation rate (SFR)
and dense molecular gas mass in the nuclei of galaxies. To do this, we
utilize the observed 850 $\mu$m luminosity as a proxy for the infrared
luminosity (\lir) and SFR, and correlate this with the observed CO
(J=3-2) luminosity. We find tentative evidence that the \lir-CO
(J=3-2) index is similar to the Kennicutt-Schmidt (KS) index ($N
\approx$ 1.5) in the central $\sim$1.7 kpc of galaxies, and flattens
to a roughly linear index when including emission from the entire
galaxy. This result may imply that the volumetric Schmidt relation is
the underlying driver behind the observed SFR-dense gas correlations,
and provides tentative confirmation for recent numerical models. While
the data exclude the possibility of a constant \lir-CO (J=3-2) index
for both galaxy nuclei and global measurements at the $\sim$80\%
confidence level, the considerable error bars cannot preclude
alternative interpretations.




\end{abstract}

\section{Introduction}
The rate at which stars form in galaxies has long been parameterized
in terms of a power-law involving the gas density \citep{sch59}.  HI
and CO observations of galaxies have shown that the star formation
rate (SFR) is related to the surface gas density via $\Sigma_{\rm SFR}
\propto \Sigma_{\rm gas}^{1.4-1.5}$ \citep{ken98a}. Theoretically, a
power-law index of $\sim$1.5 controlling the SFR is appealing. If a
constant fraction of gas is converted into stars over a free fall
time, the relation SFR $\propto \rho_{\rm gas}^{1.5}$ results.

Observations of molecular gas in galaxies have provided a more complex
view of SFR relations. Local surveys have shown that the SFR (as
traced by the \lir) is proportional to the CO (J=1-0) luminosity
\citep[$L'$;][]{gao04a} to the $\sim$1.5 power, consistent with
observed surface density SFR relations \citep{ken98a}. However, recent
observations have revealed a tight, linear relation between HCN
(J=1-0) and the \lir \ in galaxies \citep{gao04a,gao04b}. Because the
J=1-0 rotational transition of HCN has a relatively high critical
density ($n_{\rm crit} \sim$10$^{5}$ \cmthree), this has been
interpreted as a more fundamental SFR relation such that the
volumetric SFR is linearly related to the dense, star-forming
molecular gas. A linear relation between \lir \ and the luminosity
from a high critical density tracer CO (J=3-2) in a similar sample of
galaxies has provided confirming evidence for this relation and
advocated a similar interpretation \citep{nar05}.

An alternative interpretation to the linear relation between \lir
\ and dense gas tracers has been put forth from the theoretical
side. First, \citet{kru07} utilized models of individual giant
molecular clouds (GMCs) with a lognormal density distribution function
coupled with escape probability radiative transfer calculations. These
authors found that the central issue driving the observed molecular
SFR relations was the relationship between the fraction of the cloud's
gas above the critical density of the molecular transition. When, on
average, the gas density is higher than the line's critical density
(<$n_{\rm cloud}$> >> \ncrit), the bulk of the gas is thermalized in
the line, and the line luminosity increases linearly with increasing
mean gas density (<$n$>). In this scenario, a relationship between SFR
and \lmol \ is expected to have index similar to the underlying
(volumetric) Schmidt index. This is reminiscent of the observed
relation between \lir \ and CO (J=1-0) in galaxies. Alternatively,
when only a small fraction of the gas is thermalized (<$n_{\rm
  cloud}$> << \ncrit), the line luminosity will increase superlinearly
with <$n$>, and the SFR-\lmol \ relation will have index
less than that of the underlying Schmidt index (e.g. the observed
\lir-HCN J=1-0 relation).

\citet{nar08b} furthered these models of GMCs by applying 3D non-LTE
radiative transfer calculations to hydrodynamic simulations of
isolated galaxies and equal mass binary galaxy mergers.  A key finding
in these numerical models was that the observed relations are only
found when global measurements are made. Higher spatial resolution
observations of e.g. the nuclei of galaxies would probe gas in which a
larger fraction of the gas is thermalized than in unresolved
observations of the entire galaxy. In this case, <$n$>
could potentially become comparable to the critical density of the
line, and an SFR-\lmol \ index of $\sim$1.5 would be expected even for
high critical density molecular lines. In both sets of models, the
fundamental relation is the Schmidt relation with index 1.5. The
observed SFR-\lmol \ relations were simply manifestations of the
underlying Schmidt law.

 A key difference exists between the numerical models of \citet{kru07}
 and \citet{nar08b}, and the interpretations of
 \citet[][]{gao04a,gao04b}, \citet{nar05} and \citet{wu05}. The models
 find the driving relation is the volumetric Schmidt relation whereas
 the latter set of observations cite a more fundamental relation as
 that between SFR and dense molecular gas. {\it Tests that distinguish
   between these interpretations are crucial to understanding global
   SFR relations in galaxies}.  \citet{nar08b} offered a direct
 prediction from their models that observations of molecular lines
 with critical density higher than that of HCN (J=1-0) should break
 the linear trend seen between SFR and HCN (J=1-0). Specifically, for
 extremely high critical density lines, \ncrit will be even larger
 than <$n_{\rm galaxy}$> than in the case of HCN (J=1-0); In these
 cases, the relation between SFR and \lmol \ is directly predicted to
 be sublinear. Indeed, confirming evidence for this trend has been
 found by Bussmann et al. (2008; submitted), who found a sublinear
 \lir-HCN (J=3-2) index in remarkable agreement with the predictions
 of \citet{nar08b}.

An alternative generic feature of both the \citet{kru07} and
\citet{nar08b} simulations which may serve as a test of the models is
a superlinear SFR-\lmol \ relation for high critical density tracers
when <$n_{\rm galaxy}$> is sufficiently high. One
potential manifestation of this is a break in the linear SFR-HCN
(J=1-0) relation in systems with extremely high infrared luminosity
(e.g. hyper-luminous infrared galaxies; \lir >
10$^{13}$\lsunend). Here, the models predict a steepening in the
SFR-\lmol \ index toward $N$=1.5 \citep{kru07}. Tentative evidence for
this may have been found by \citet{gao07} in high redshift systems,
though the potential contribution of active galactic nuclei (AGN) to
the \lir \ may drive a similar signature.

In order to avoid the muddying effects of central AGN, a potential
alternative to this test is to observe the SFR-\lmol \ relation in
galactic nuclei, where the local mean gas density may be higher than
the globally averaged mean gas density. In this {\it Letter}, we
utilize literature data in order to examine the relationship between
SFR (traced by the 850$\mu$m flux) and dense molecular gas (traced by
\lcojthree). The aim is to help distinguish between the competing
interpretations of the linear SFR-dense gas relations in galaxies.

\section{Literature Data}
\label{sec:obs}


In order to investigate the relation between SFR and dense molecular
gas over a variety of physical spatial extents in galaxies, we require
a beam matched set of observations in both high critical density
molecular line and SFR tracer with a sufficient number of galaxies.
While matching the highest resolution and sensitivity infrared data of
nearby galaxies \citep[e.g SINGS; ][]{ken03} with high spatial
resolution observations of a high critical density tracer
\citep[e.g. ][]{kri07} would be ideal, few galaxies exist at the
intersection of such surveys. Most other traditional infrared surveys
are insufficient as they typically report global measurements, rather
than higher spatial resolution data. Further complications arise on
the dense gas tracer side. Because of the prodigious observing time
necessary to map large numbers of galaxies with an interferometer,
most dense gas surveys are done with a single dish and report on lines
at $\sim$mm wavelengths \citep[e.g.][]{baa08}. In order to obtain the
highest spatial resolution possible from single dish surveys,
higher-frequency observations (e.g. $\lambda$<1 mm) must be employed.

The largest beam-matched sample of a dense gas tracer and \lir
\ tracer is the JCMT CO (J=3-2) survey of \citet{yao03} and the SCUBA
Local Universe Galaxy Survey \citep{dun00}. \citet{yao03} used the
850$\mu$m images of these galaxies to scale the total \lir \ to match
the $\sim$15$\arcsec$ beam of the CO (J=3-2) observations.  As such,
while the number statistics are relatively small, to our knowledge
this sample comprises the largest available for this type of analysis.
Moreover, using CO (J=3-2) as a tracer of dense molecular gas has the
attractive quality that both global measurements \citep{nar05} and
simulated unresolved observations from simulations \citep{nar08b} have
shown a linear relation between \lir \ and CO (J=3-2) luminosity in
galaxies.

In order to avoid single galaxies at the extrema of the \lir \ range
artificially biasing the fits, we impose nominal luminosity cuts in
the sample, considering galaxies in the range \lir=[10$^9$,
  2$\times$10$^{11}$ \lsunend]. This excludes galaxies which may be
interacting, thus allowing us to use galaxy distance as a proxy for
mean density (\S~\ref{sec:results}). This results in a total sample
size of 40 galaxies, excluding only 4 galaxies from the parent sample.

\section{Results}
\label{sec:results}


\begin{figure}
\includegraphics[scale=0.5]{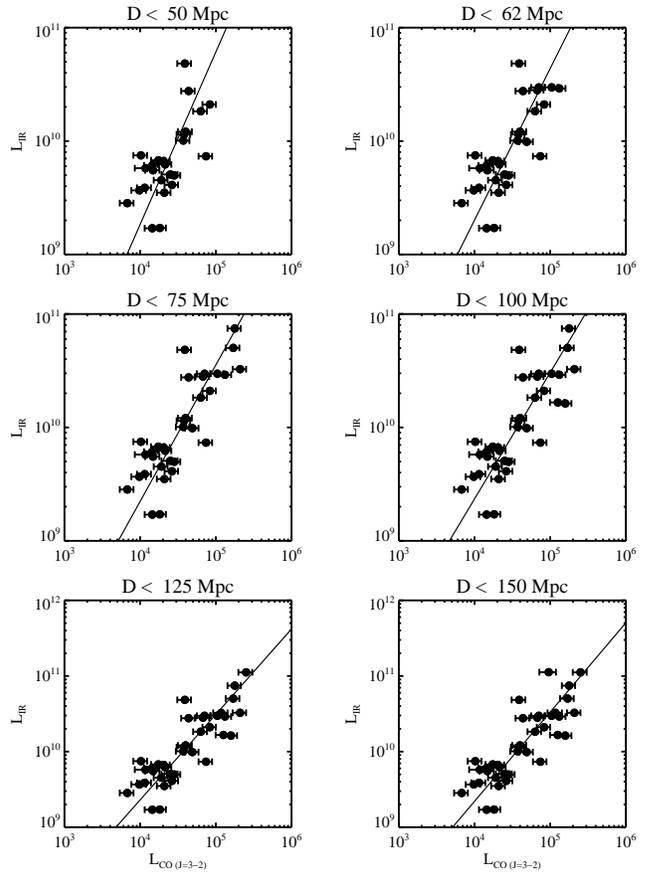}
\caption{The \lir-\lcojthree \ relation for galaxies in each of our
  distance bins. The slope is seen to flatten as observations probe
  more global measurements.\label{figure:lirvlco32}}
\end{figure}

\begin{figure}
\includegraphics[scale=0.375,angle=90]{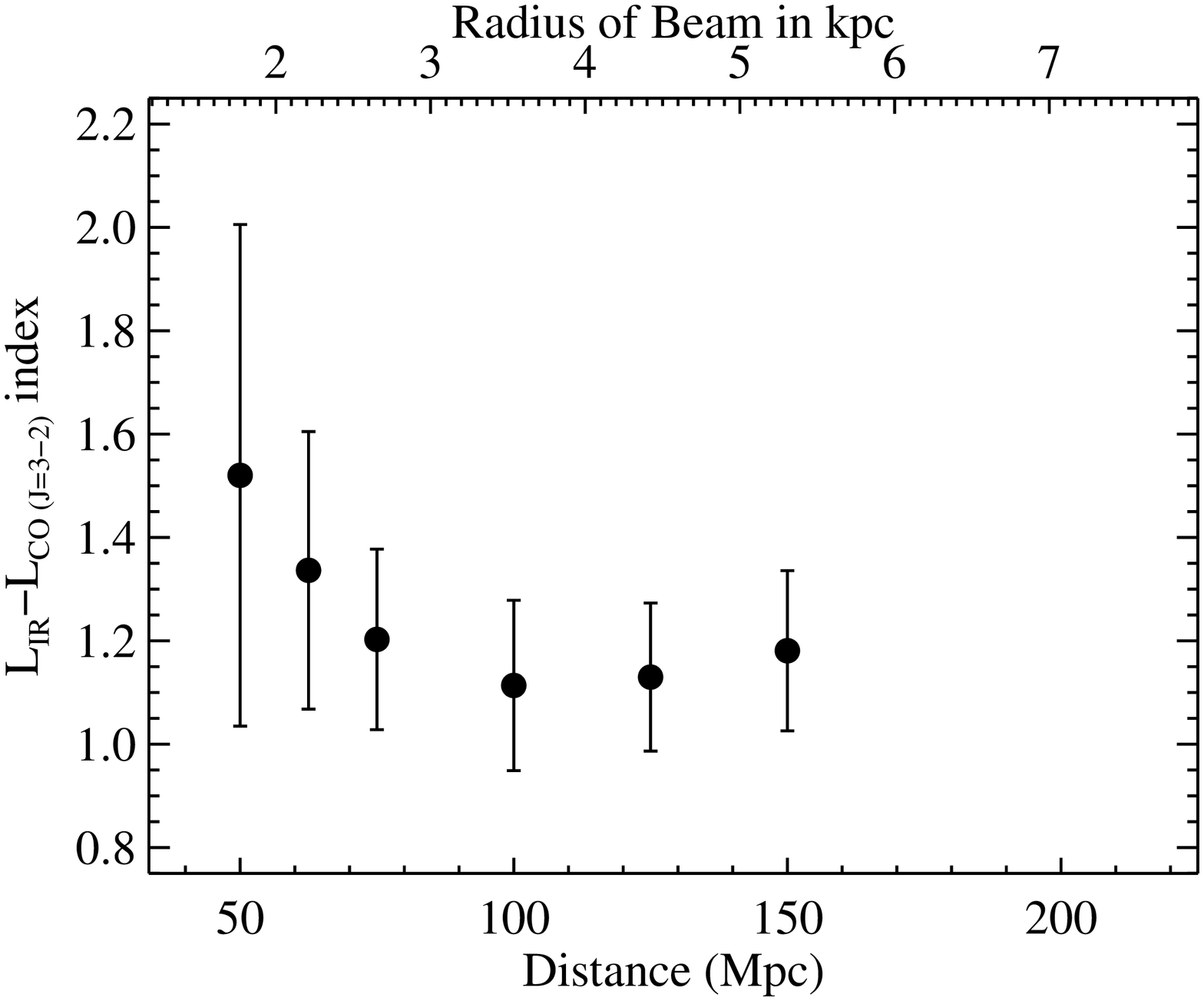}
\caption{The \lir-CO (J=3-2) index as a function of limiting galaxy
  distance. The physical scale covered by the beam is on the top
  axis. Observations of the nearby galaxies probe the nuclear region
  of the galaxies. In this high <$n$> regime, the \lir-CO (J=3-2)
  index tentatively lies near the volumetric Schmidt index, also
  consistent with recent modeling efforts by \citet{kru07} and
  \citet{nar08b}. As observations become unresolved and more low
  density gas is folded into the beam, the index tends toward
  unity. \label{figure:lirvlco32_index}}
\end{figure}

We have binned our sample of galaxies into bins in distance. Because
we are considering observations of a fixed beamsize, this is
equivalent to binning the galaxies in terms of physical radial extent
from the nucleus probed by the observations. 
The number of galaxies rises sharply to $\sim$25 when considering
galaxies with a maximum distance of $\sim$50 Mpc. We thus utilize 50
Mpc as the first distance bin in order to maximize the statistical
significance of the result.

In Figure~\ref{figure:lirvlco32}, we show the \lir-\lcojthree
\ relation for each distance bin. In
Figure~\ref{figure:lirvlco32_index}, we plot the best fit log(\lir)
-log(\lcojthree) \citep[L';][]{gao04a} slope as a function of limiting
distance.  On the top axis we label the physical extent of the
beamsize.
The fits were done utilizing the publically available Monte Carlo
Markov Chain fitting packages of \citet{kel07}. The Bayesian routines
assume the intrinsic distribution can be well approximated via a
mixture of Gaussians, and returns the posterior probability
distribution function (pdf) of potential slopes to the fit. Unlike
standard linear regression techniques, the fit is not done with the
assumption that the abscissa values are known exactly while
measurement errors exist only for the ordinate. Rather, errors are
allowed in both axes. This is important as errors are only reported
for the literature CO measurements \citep{yao03}, and to conform with
standard literature fits, we are required to fit \lir \ as a function
of the CO luminosity. The points in
Figure~\ref{figure:lirvlco32_index} are the median of the returned
distribution of slopes, and the error bars denote the standard
deviation in the pdf.

The slopes in the \lir-\lcojthree \ relation tentatively show a
similar trend to what would be expected from the models of
\citet{kru07} and \citet{nar08b}. Specifically, in the regions of high
<$n$> where the bulk of the gas may be thermalized, the SFR-\lmol
\ relation for high critical density tracers tends toward 1.5
(e.g. the points at D$\approx$50 Mpc in
Figure~\ref{figure:lirvlco32_index}; this corresponds to the central
$\sim$1.7 kpc for most of the galaxies in this bin). When considering
more unresolved, global observations of galaxies which fold in
significant amounts of diffuse gas (thus lowering the effective <$n$>
probed by the observations), the mean density of the galaxy drops
below the critical density, and the observed SFR-\lmol \ index is less
than that of the underlying volumetric Schmidt index. Here, we see
that the more unresolved observations approach a slope of unity,
consistent with the linear slope found between \lir \ and CO (J=3-2)
luminosity in a nearly identical sample of galaxies by
\citet{nar05}. We note that the slope is still moderately superlinear
in these bins, though consistent with the range of results expected by
the \citet{nar08b} models. This occurs because the maximum distance
bins still contain the relatively nearby (D<50 Mpc) galaxies, which
increases the \lir-\lcojthree \ index. Ideally one would like to
exclude the nearby galaxies from the most distant bins, but low number
statistics prevent this experiment.

\section{Discussion}
\label{sec:discussion}
The trends seen in Figure~\ref{figure:lirvlco32_index} may be
consistent with a picture in which the observed SFR-\lmol \ relations
are driven globally by the relationship between <$n_{\rm gas}$> and
\ncrit. In contrast, if the linear relation between \lir \ and HCN
(J=1-0) or CO (J=3-2) were indicative of a more fundamental SFR
relation in galaxies in terms of dense molecular gas, one would expect
the relation to remain linear even in the nuclei of galaxies. With a
potentially model distinguishing result such as this one, obvious
questions regarding its robustness arise.

\begin{figure}
\includegraphics[angle=90,scale=0.375]{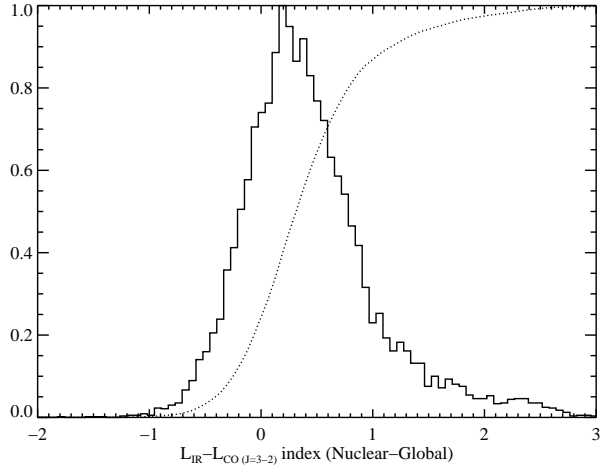}
\caption{Solid Line: Normalized difference in PDFs of \lir-\lcojthree
  \ indices for lowest distance bin and largest distance bin. Dotted
  line: Cumulative distribution function for shown PDF. This can be
  interpreted as the PDF of the \lir-\lcojthree \ indices for the
  nuclear and global observations as being the
  same.  \label{figure:cumdist}}
\end{figure}

First, we caution that the error bars presented in the lowest distance
bins in Figure~\ref{figure:lirvlco32_index} are rather large. This
owes to the small sample sizes of these bins ($\sim$25 galaxies for
the lowest distance bins). It is possible to investigate what the
probability is that the same \lir-\lcojthree \ index properly
characterizes both the lowest and highest distance bins. To do this,
we remind the reader that each point (and associated error bars) in
Figure~\ref{figure:lirvlco32_index} is a pdf for potential
\lir-\lcojthree \ indices at each distance bin. In
Figure~\ref{figure:cumdist}, we plot the (normalized) difference in
the pdf's from the highest and lowest distance bins in
Figure~\ref{figure:lirvlco32_index}. We additionally show the
cumulative distribution function in the same plot. A value of 0 in the
pdf difference is expected at points when the same index characterizes
both the high and low distance bins. As can be seen, the nuclear
\lir-\lcojthree \ index is systematically weighted toward larger
numbers than the global value. That said, the error bars in the
nuclear distance bin are not insignificant. The probability that the
\lir-\lcojthree \ indices from the nuclear and global distance bins
are the same $\pm$0.25(0.5) is $\sim$17(33)\%. Thus, while the
tentative trends seen in Figure~\ref{figure:lirvlco32_index} are
indeed probable, they are by no means robust. In this sense, surveys
to increase the number statistics will be required to confirm/refute
this potential result.


Second, we can inquire as to the validity of \lir \ as an SFR
indicator. In particular, the \lir \ from galaxies has the potential
to be contaminated by central AGN. That said, the galaxies in the
sample presented here are relatively low luminosity (global \lir \ <
10$^{12}$ \lsunend). Both observational studies \citep{tra01} and
theoretical works \citep{cha07a} suggest that galaxies with the \lir \
range investigated here have their IR luminosity largely powered by
star formation.

\begin{figure}
\includegraphics[angle=90,scale=0.375]{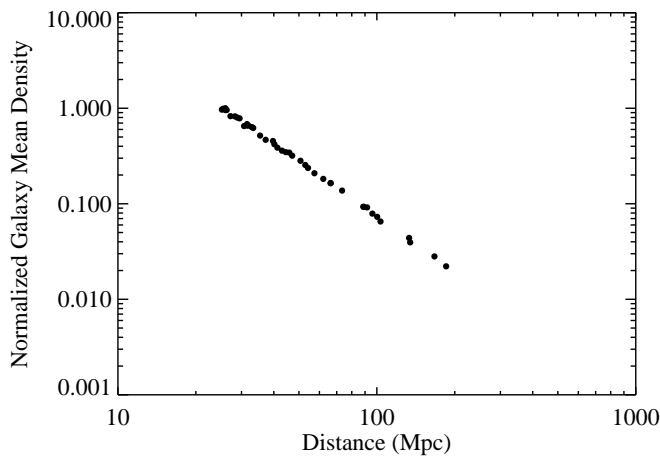}
\caption{The normalized <$n$> within the CO (J=3-2) beam as a function
  of galaxy distance. The <$n$> within the beam drops as a function of
  distance, showing that the more unresolved observations are probing
  lower <$n$> gas. This implies that the relationship seen in
  Figure~\ref{figure:lirvlco32_index} owes to a dropping <$n$> in the
  farthest objects. \label{figure:meandens}}
\end{figure}

Third, we can question whether or not the trend in the \lir-CO (J=3-2)
index seen with distance is equivalent to a trend in <$n$>. To do
this, in Figure~\ref{figure:meandens}, we plot the normalized <$n$> of
the galaxy versus the galaxy distance.  For the density determination,
the volume is derived by assuming the gas resides in a disk with
radius equivalent to the physical extent probed by the beam at that
distance, and a constant scale height in each galaxy. The molecular
gas mass is taken from the CO (J=1-0) luminosity, utilizing a CO-\htwo
\ conversion factor appropriate for this sample of galaxies
\citep{yao03}.  The CO (J=1-0) observations are beam matched to the CO
(J=3-2) observations presented here, so the molecular gas mass is
indeed the gas mass within the radial extent of the CO (J=3-2)
beam. The mean density is normalized to account for a variety of
uncertain parameters (e.g. disk scale height, molecular gas volume
filling factor, density profile of GMCs). Because we are only
concerned with the trend in <$n_{\rm galaxy}$> with distance,
normalizing the density has no
consequence. Figure~\ref{figure:meandens} shows that the mean density
probed by the molecular line observations drops as we consider
galaxies successively farther away. Alternatively said, unresolved
observations of galaxies probe lower <$n$> than observations toward
galactic nuclei. This implies that the tentative trend in \lir-CO
(J=3-2) index seen in Figure~\ref{figure:lirvlco32_index} owes to the
relationship between the mean density of the gas in the beam and
\ncrit \ of the emission line.

If larger samples verify the results of
Figure~\ref{figure:lirvlco32_index}, then these observed trends may
have several implications. The models of \citet{kru07} and
\citet{nar08b} which predict an SFR-\lmol \ index of 1.5 when <$n$> is
high are predicated on an underlying relation controlling the SFR: SFR
$\propto$ $n^{1.5}$. The results presented here tentatively support a
scenario in which the SFR can be described by such a power-law (rather
than a linear relation), though we reiterate that the error bars in
this study are substantial.

In principle, we could perform the same experiment shown in
Figure~\ref{figure:lirvlco32_index} with the CO (J=1-0)
transition. Indeed, Nobeyama 45m observations of CO (J=1-0) in many of
the same galaxies exist, thus providing the opportunity for a
beam-matched comparison to this dataset \citep{yao03}. However, the
error bars on these observations are sufficiently large that the
slopes on the \lir-\lcojone \ indices for the lowest distance bins (D
< 100 Mpc) are consistent with numerous interpretations (including no
correlation at all).

The lack of clean data in low \ncrit \ transitions does not preclude
us from conjecturing what results one might expect in a similar
experiment with CO (J=1-0). Naively, one might imagine the \lir-CO
(J=1-0) index to be $\sim$1.5 for all distance bins owing to the
relatively low critical density of this ground state
transition. However, the more likely scenario are results similar to
the trend seen in Figure~\ref{figure:lirvlco32_index} - a slope of
$\sim$1.5 for the lowest distance bins, and a flattening to a slope of
$\sim$1. The reason for this is global observations of galaxies in
this relatively low \lir \ range have a linear \lir-CO (J=1-0)
relation \citep{gao04a,gao04b}. It is only when considering galaxies
with \lir $\ga$10$^{11}$ \lsun in the sample that unresolved
observations result in an \lir-\lcojone \ index of 1.5. In the
interpretation of \citet{kru07} and \citet{nar08b}, this would result
from lower luminosity galaxies having most of their gas subthermal in
this transition.

Finally, we emphasize that though the results presented here advocate
an interpretation of the linear \lir-dense gas relations consistent
with \citet{kru07} and \citet{nar08b}, this is not to say that some
high critical density tracers such as HCN (J=1-0) are not applicable
as SFR tracers. By virtue of their linearity with \lir, insomuch that
the \lir \ can be used as an adequate proxy for the SFR, so can global
measurements of HCN (J=1-0) and CO (J=3-2).

\section{Conclusions and Ways Forward}
\label{sec:conclusions}

We have studied the relationship between the SFR and dense molecular
content in the nuclei of galaxies by utilizing the observed \lir \ as
a proxy for the SFR, and CO (J=3-2) as a proxy for dense molecular
gas. We find tentative evidence that the \lir-CO (J=3-2) index is
superlinear in galactic nuclei with value $N \sim$ 1.5, while
flattening as the observations measure more global conditions. 

The error bars in this potential trend are considerable however, and
that the results here serve as motivation for future observational
experiments investigating the \lir-dense gas relation in a variety of
environments. One potential way to increase the sample size would be
to obtain additional beam-matched sub-mm and CO (J=3-2) observations
of the nuclei of nearby galaxies with 15 m sub-mm telescopes (e.g. the
JCMT or ASTE). Alternatively, high spatial resolution observations of
nearby galaxies with ancillary IR maps (e.g. the SINGS sample), would
provide a rich dataset with \lir \ and dense gas measurements for
numerous regions with varying physical conditions in individual
galaxies. Additional data to constrain the gas density (e.g. CO J=1-0)
would allow the \lir-\lcojthree\ index to be studied truly as a
function of gas density, as opposed to galaxy distance. These
experiments will help to confirm/refute the potential results of
Figure~\ref{figure:lirvlco32_index}.

\acknowledgements We thank Lihong Yao and Brandon Kelly for helpful
conversations. Support for this work was provided in part by the Keck
Foundation.

\end{document}